\documentstyle[12pt]{article}
\begin{document}

\title{Detecting Scalar Higgs Bosons in the Next-to-Minimal
Supersymmetric Standard Model at Future Linear Colliders}
\author{B. R. Kim\thanks{E-mail: kim@physik.rwth-aachen.de
        (telephone +49-241-80-7239)}    
        ~~and G. Kreyerhoff
        \\III. Phys. Inst. A, Physikzentrum, RWTH Aachen \\
          Aachen, 52056, Germany
\\
        \\S. K. Oh
        \\Department of Physics, Kon-Kuk University
        \\ Seoul 143-701, Korea
\\
\\
\\}
\date{}

\maketitle
\thispagestyle{empty}

\begin{abstract}
We analyze the possibility of detecting one of the three scalar Higgs
bosons in the next-to-minimal supersymmetric model at the future
$e^+e^-$ linear collider, by examining their productions via the
Higgsstrahlung process.
The production cross sections of the three scalar Higgs bosons
in $e^+e^-$ collisions are evaluated for the proposed c.m. energies of
the future $e^+e^-$ colliders, for the whole space of relevant 
parameters.
We find that at least one of the three production cross sections
is not smaller than 16 fb, 4 fb, and 1 fb for ${\sqrt s} = 500$ GeV,
1000 GeV, and 2000 GeV, respectively.
Those numbers indicate that at least one of the three scalar Higgs
bosons in the next-to-minimal supersymmetric model may be detected
at the future $e^+e^-$ linear colliders via
the Higgsstrahlung process.
\end{abstract}

\vfil
\eject


The scalar Higgs bosons constitute an essential part of the standard 
model
and the extended models therefrom [1]. To search for and discover them~
is thus one of the most important tasks of both theoretical and
experimental particle physicists.
As is well known, there is just one scalar Higgs boson, $S$, in the
standard model, whereas there are multiples of them in the extended 
models.
For example, the minimal supersymmetric standard model (MSSM) [2] has
two scalar Higgs bosons, $S_1$ and $S_2$, while the next-to-minimal
supersymmetric standard model (NMSSM) [3] has three of
them: $S_1$, $S_2$, and $S_3$.

The MSSM has just two guage doublet Higgs superfields, ${\cal H}_1$
and ${\cal H}_2$, and the NMSSM is a minimal extension of the MSSM,
by introducing an additional gauge singlet Higgs superfield
${\cal N} = (N, \psi_N, F_N)$ to the Higgs sector of the MSSM.
Here, $N$ is a Higgs singlet, $\psi_N$ is a singlet higgsino, and
$F_N$ is an auxiliary field.
As mentioned elsewhere [4], this singlet superfield can provide an
economical way to avoid the so called $\mu$-parameter problem in the 
MSSM.
Thus, while the MSSM introduces the dimensional $\mu$-parameter by hand, 
the NMSSM can generate it in terms of the singlet superfield ${\cal N}$. 

In the NMSSM, the relevant superpotential is given by
\begin{equation}
      {\cal W} = \lambda  {\cal H}^T_1 \epsilon {\cal H}_{2} {\cal N}
                - \frac{1}{3}k {\cal N}^3 + \cdots
\end{equation}
where $\lambda$ and $k$ are dimensionless parameters.
We assume that they are real.
The relevant soft-breaking part of the Higgs sector is accordingly given 
by
\begin{equation}
        V_{\rm soft} = -\lambda A_{\lambda} H^{T}_{1} \epsilon H_{2}N~
                - \frac{1}{3} kA_{k}N^{3} + {\rm h.c}. + \cdots
\end{equation}
where $A_{\lambda}$ and $A_k$ are soft-breaking parameters having
mass dimension.
As the vacuum expectation value $\langle N \rangle_{0} = x$ is 
developed,
$\lambda x$ emerges to correspond to the $\mu$-parameter in the MSSM
superpotential ${\cal W} = \mu{\cal H}^T_1\epsilon{\cal H}_2 + =
\cdots$.
Note that the Peccei-Quinn symmetry is recovered if $k =0$.
Thus, $\lambda k \neq 0$.
The two Higgs doublets $H_1$ and $H_2$ give masses to the up-quark 
sector
and the down-quark sector, when they develop the vacuum
expectation values, $v_1$ and $v_2$, respectively.

A side effect of the introduction of the new superfield is that
it enriches the particle spectra of the NMSSM as well as introduces more 
parameters than the MSSM.
While the charged sector of the NMSSM is the same as that of the MSSM,
the neutral sector of the NMSSM has an extra neutralino,
an extra scalar Higgs and an extra pseudoscalar Higgs boson.
The number of parameters in the NMSSM is also larger than in the MSSM.
For example, while one needs just two parameters in order to determine
the Higgs boson masses in the MSSM, which can be chosen as $\mu$ and
$\tan \beta = v_1/v_2$, six parameters are introduced for the same
purpose in the NMSSM:
$\lambda$, $x$, $k$, $A_{\lambda}$, $A_k$, and $\tan \beta$.

The present experiments at LEP set lower bounds on the mass of the~
lightest scalar Higgs boson.
The standard scalar Higgs boson may not be lighter than 65 GeV,
while $S_1$ in the MSSM may not have a mass below 45 GeV as it has not
been detected yet [5].
For the scalar Higgs bosons in the NMSSM, it is shown
that there are parameter regions where the lightest scalar Higgs
boson $S_1$ may have a small mass without contradicting to the
present experimental data [6].

Our main purpose in this article is to show that at least
one of the scalar Higgs bosons in the NMSSM can be detected at the 
future
$e^+e^-$ Linear Colliders with ${\sqrt s}$ = 500, 1000, or 2000 GeV.
This is done by calculating the minima of the production cross sections
of the three scalar Higgs bosons in $e^+e^-$ collisions.
We find that there are parameter regions in the NMSSM where at least
one of the minima of the production cross sections are larger than
the proposed discovery limit of the future $e^+e^-$ colliders,
which implies that the future $e^+e^-$ colliders are capable of
detecting at least one of the three scalar Higgs bosons in the NMSSM.
A preliminary result of our analysis is reported elsewhere [7].


A remarkable result of the MSSM is that there is an upper bound on
the mass of the lighter scalar Higgs boson given by
\[
        m_{S_1} \le m_Z \cos 2\beta
\]
at the tree level. This is due to the fact that all quartic terms have
gauge coupling constants.
In case of the NMSSM with one Higgs singlet,
there are quartic terms with couplings other than the gauge couplings.
Thus, there is no simple upper bound on the mass of scalar Higgs boson
as in the MSSM.
The tree-level upper bound on $m_{S_1}$ is given as [8]
\begin{equation}
         m^2_{S_1} \le (m^{\rm max}_{S_1})^2
                = m^2_Z\; (\cos^2 2\beta
           + {2\lambda^2 \cos^2 \theta_W \over g^2_2}\sin^2 2\beta )
\end{equation}
It approaches to the MSSM relation $m^2_{S_1} < m^2_Z$ for
$\tan \beta \rightarrow \infty$ or $\lambda \rightarrow 0$.

The above relation may be cast into a simple form, by noticing that
the coefficient of $\lambda$ above, $2\cos^2\theta_W /g^2_2$, is very
close to the neutrino-$Z$ coupling constant
$g_{\nu\nu Z} = g_2/2\cos\theta_W$.
Thus, denoting $\lambda_0 = g_2/{\sqrt2}\cos\theta_W = 
{\sqrt2}g_{\nu\nu Z}$,
one has $\lambda_0 (m_Z) = 0.52$ for $g_{\nu\nu Z} (m_Z) = 0.3714$.
With this value, the upper bounds on the $S_1$ mass at tree level is:
$m_{S_1} \leq m_Z$ for $\lambda \leq \lambda_0 = 0.52$ and
$m_{S_1} \leq {\lambda \over \lambda_0}m_Z = 1.92\, \lambda\, m_Z$
for $\lambda > \lambda_0$ [8].

The dependence of $m_{S_1}$ on $\lambda$ comes from the fact that there
is a quartic term with the coupling constant $\lambda$ in the NMSSM.
It turns out that the upper bound on $\lambda$ may be relevant for
$m_{S_1}$. An effective way of determining the upper bound on $\lambda$
is the renormalization group (RG) equation analysis [9,10].

The 1-loop RG equations involving $\lambda$ read as
\begin{eqnarray}
   {d\lambda \over dt} &=& {1\over 8\pi^2}(k^2 + 2\lambda^2
               +{3\over2}h^2_t -{3\over2}g^2_2 - {1\over2}g^2_1)\lambda~
    \cr \cr
   {dk \over dt} &=& {3\over 8\pi^2}(k^2 +\lambda^2)k \cr \cr
   {dh_t \over dt} &=& {1\over 8\pi^2}({1\over2}\lambda^2
               +3h^2_t -{8\over3}g^2_3 -{3\over2}g^2_2
               -{13\over 18}g^2_1) h_t  \\ \cr
   {d g_1 \over dt} &=& {11\over 16\pi^2}g^3_1 \cr \cr
   {d g_2 \over dt} &=& {1\over 16\pi^2} g^3_2 \cr \cr
   {d g_3 \over dt} &=& -{3\over 16\pi^2} g^3_3 \cr \nonumber
\end{eqnarray}
where $t = \ln \mu$, $\mu$ being here the renormalization scale.
Numerical integration of the above RG equations by demanding the
existence of no Landau poles up to the GUT scale yield the upper
bound of $\lambda$ at the electroweak scale.
We plot $\lambda_{\rm max}$ as function of $\tan\beta$ for some values
of $k$, for $m_t = 175$ GeV in Fig. 1 and for $m_t = 190$ GeV in 
Fig. 2.

We find that $0.64 \le \lambda_{\rm max} \le 0.74$
for $175 \le m_t \mbox{ (GeV)} \le 190$.
The upper bound on $\lambda$ is roughly independent of $\tan\beta$
for $\tan\beta \ge 3$.
Also, the two figures show that $\lambda_{\rm max}$ decreases with
increasing $k$. The upper bound on $k$ is about 0.7.
Moreover, the lower bound on $\tan\beta$ is set at the electroweak 
scale.
We have $\tan\beta \ge 1.24$ for $m_t = 175$ GeV and
$\tan\beta \ge 2.6$ for $m_t = 190$ GeV.

Using $0.64 \le \lambda_{\rm max} \le 0.74$, we obtain
the tree-level upper bound on $m_{S_1}$ as
\begin{equation}
        113 \le m^{\rm max}_{S_1} \mbox{ (GeV)} \le 131
\end{equation}
Our results may well be compared with previous analyses.
For example, the upper limit of $\lambda$ has been estimated to be
about 0.87 [9] for $h_t \geq 0.5$, where $h_t$ is the top
Yukawa coupling, which yields $m_{S_1} \leq 151$ GeV.
By similar analysis of RG equations, the upper limit of $\lambda$ was
obtained as function of $h_t$, thus yielding an upper bound
on $m_{S_1}$ as 140 GeV [10].

However, the above tree-level upper bound does not contain radiative~
corrections to the mass matrices of Higgs bosons.
As in the case of the MSSM, the contributions from the radiative
corrections is found to change the tree-level upper bound considerably.
Several groups have calculated the higher-order contributions to the
mass matrices of Higgs bosons and determined the corrected upper
bound on $m_{S_1}$ [8,11,12].

We employ the effective potential method of Coleman and Weinberg [13]
to evaluate the higher-order contributions at the 1-loop level.
The 1-loop effective potential can then be written as [11]
\begin{equation}
       V_1 (Q) = V_0 (Q) + \Delta V_1 (Q)
\end{equation}
where $V_0(Q)$ is the tree-level potential evaluated with couplings
renormalized at the scale $Q$, and
\begin{equation}
    \Delta V_1 (Q) = {1\over 64\pi^2}{\rm STr}\,
                     {\cal M}^4 (\ln{{\cal M}^2\over Q^2} - {3\over2})
\end{equation}
is the 1-loop contribution.
Here, ${\cal M}^2$ is the field-dependent generalized mass matrix
containing mass terms for all the particles in the NMSSM.

The mass matrix $M^S$ of the scalar Higgs bosons at 1-loop level is,
to a good approximation [11], given by the second derivative of $V_1(Q)$ 
with respect to the Higgs fields.
Taking only top quark and stop quark contributions into account,
and assuming the degeneracy of the left and the right stop quark mass,
we find that the higher-order contributions change the relevant elements 
of the tree-level mass matrix as [8]
\begin{eqnarray}
       & &M^S_{11} \rightarrow M^S_{11} + D_{11} \cr
       & &M^S_{12} \rightarrow M^S_{12} + D_{12} \\
       & &M^S_{22} \rightarrow M^S_{22} + D_{22} \nonumber
\end{eqnarray}
where
\begin{eqnarray}
      D_{11} &=& -{1\over16\pi^2}\left({\lambda x A_T \over 
v_2}\right)^2
                  \left({m_t \over m_{\tilde t}}\right)^4  \cr \cr
      D_{12} &=& {3\over 8\pi^2}\lambda x A_T~
                  \left({m^2_t \over m_{\tilde t} v_2}\right)^2
                  \left(1+ {1\over6}{A_t A_T \over m^2_{\tilde 
t}}\right)
                   \\ \cr
      D_{22} &=& {3\over 8\pi^2}\left({m^2_t \over v_2}\right)^2
                  \left[2\ln\left({m_{\tilde t}\over m_t}\right)^2
                        -{2A_t A_T \over m^2_{\tilde t}}
                        -{1\over6}{A^2_t A^2_T \over M^4_{\tilde t}}
                                  \right] \cr \nonumber
\end{eqnarray} ~
with $A_T = -A_t +\lambda x \cot\beta$.
The $b$-quark contributions are not included as they are negligibly
smaller than the $t$-quark contributions.
Those higher-order contributions render the upper bound on $m_{S_1}$ as
\begin{equation}
        (m^{\rm max}_{S_1})^2 \rightarrow
        (m^{\rm max, 1-loop}_{S_1})^2 =       (m^{\rm max}_{S_1})^2
        + D_{11} \cos^2\beta + D_{12} \sin 2\beta + D_{22} \sin^2\beta
\end{equation}

Incorporating the upper bound of the parameter $\lambda$ from the
RGE analysis, we determine the upper bound on $m_{S_1}$ in the
parameter regions of the NMSSM for
$250 \le x \mbox{ (GeV)} \le 1000$, $250 \le A_t \mbox{ (GeV)} \le =
1000$,
and $250 \le m_{\tilde t} \mbox{ (GeV)} \le 1000$,
where $x$, $A_t$, and $m_{\tilde t}$ are respectively the
vev of the Higgs singlet, the soft breaking coefficient of top sector,
and stop quark mass, and for $2 \le \tan\beta \le 20$,
for $175 \le m_t \mbox{ (GeV)} \le 190$.
Our result is:
\begin{equation}
        120 \le m^{\rm max, 1-loop}_{S_1} \mbox{ (GeV)} \le 156
\end{equation}
for the lightest scalar Higgs boson in the NMSSM at the 1-loop level 
[8].

Also, in terms of $m^{\rm max}_{S_1}$, it can be shown that the
upper bounds on $m_{S_2}$ and $m_{S_3}$ are given as
\begin{eqnarray}
        & & m^2_{S_2} \le (m^{\rm max}_{S_2})^2
                = {(m^{\rm max}_{S_1})^2 -R^2_1 m^2_{S_1}
                        \over 1 - R^2_1} \cr
        & & m^2_{S_3} \le (m^{\rm max}_{S_3})^2
                = {(m^{\rm max}_{S_1})^2 -(R^2_1 +R^2_2) m^2_{S_1}
                        \over 1 - (R^2_1 +R^2_2)} \nonumber
\end{eqnarray}
where $R_1 = U_{11}\cos\beta + U_{12} \sin\beta$ and
$R_2 = U_{21}\cos\beta + U_{22} \sin\beta$, $U_{ij}$ being
the $3\times 3$ orthogonal matrix which diagonalizes the mass
matrix $M^S$ of the scalar Higgs bosons.
Clearly, $R_1$ and $R_2$ are complicated functions of the relevant
parameters of the NMSSM.
Note moreover that they satisfy the unitarity condition
$0 \le R^2_1 +R^2_2 \le 1$.


That the upper bound on $m_{S_1}$ in the NMSSM at 1-loop level is
between 120 and 156 GeV suggests that
the accessible region of the parameter space at LEP 1
with ${\sqrt s} = m_Z$ might be very small.
Actually, we have shown that the existing LEP 1 data do not exclude
the existence of $S_1$ with $m_{S_1} = 0$ GeV [6].

For the future colliders with ${\sqrt s} = 500$, 1000, or 2000 GeV,
the situation is quite different.
In this case, the production of the lightest scalar Higgs boson
via the Higgsstrahlung process, $e^+e^- \rightarrow S_1 Z$ where
$Z$ decays further into a pair of fermion and antifermion, with
real $Z$ and real $S_1$, is always possible, because the collider
energy ${\sqrt s}$ is much larger than $E_T = m_Z + m_{S_1}$:
\[
        212 \le E^{\rm max}_T = m_Z + m^{\rm max}_{S_1}
        \mbox{ (GeV)} \le 248
\]
Thus, $E_T$ plays a kind of threshold energy and is an important
quantity of our model.
To obtain informations about how far the model could be tested, then,
it would be helpful to derive a lower bound on the production cross
section of the Higgsstrahlung process, which can be expressed as
a function of the collider energy only.

At the proposed center of mass energies of 500, 1000, or 2000 GeV for
the future $e^+e^-$ linear colliders, the question is therefore not
whether it is kinematically possible to produce $S_1$ but whether
the production rate is large enough for $S_1$ to be detected.
If the production rate of $S_1$ be small, one should then examine as
a next step if that of $S_2$ or $S_3$ is large enough.
In order to by systematic, we consider all of the productions of
$S_1$, $S_2$, and $S_3$ via the Higgsstrahlung process.

The cross sections for the productions of the three scalar Higgs bosons
via the Higgsstrahlung process can be expressed as
\begin{eqnarray}
        \sigma_1 (R_1, R_2, m_{S_1}) &=& \sigma_{SM}(m_{S_1}) R^2_1  \cr
        \sigma_2 (R_1, R_2, m_{S_2}) &=& \sigma_{SM}(m_{S_2}) R^2_2  \\
        \sigma_3 (R_1, R_2, m_{S_3})~
                &=& \sigma_{SM}(m_{S_3}) (1 -R^2_1 -R^2_2) \nonumber
\end{eqnarray}
where $\sigma_{SM}(m)$ is the cross section in the standard model
for the production of the Higgs boson of mass $m$ via the Higgsstrahlung 
process.
Note that the production cross sections are also functions of ${\sqrt s}$ 
of the collider, though not explicitly shown.
A useful observation is that
$\sigma_i (m^{\rm max}_{S_i}) \le \sigma_i (m_{S_i})$ which allows one 
to derive the parameter-independent lower bound on $\sigma_i$, as we will
see shortly.

Now, we calculate for given $R_1$ and $R_2$ the cross section
$\sigma_1 (R_1, R_2, m_{S_1})$ for $S_1$ production via the 
Higgsstrahlung
process. We repeat the calculations for $S_2$ and $S_3$ productions
to obtain $\sigma_2 (R_1, R_2, m_{S_2})$
and $\sigma_3 (R_1, R_2, m_{S_3})$.
Because the production cross sections decrease as the produced Higgs
boson mass increases, one has
\begin{eqnarray}
        & & \sigma_2 (R_1, R_2, m_{S_2})
        \ge \sigma_2 (R_1, R_2, m^{\rm max}_{S_2}) \cr
        & & \sigma_3 (R_1, R_2, m_{S_3})
        \ge \sigma_3 (R_1, R_2, m^{\rm max}_{S_3})
\end{eqnarray}

The second step is to calculate, by allowing $m_{S_1}$ to vary
from its minimum to its maximum value, for given $R_1$ and $R_2$,
the set of the three production cross sections
$\sigma_1 (R_1, R_2, m_{S_1})$, $\sigma_2 (R_1, R_2, m^{\rm max}_{S_2})$,
and $\sigma_3 (R_1, R_2, m^{\rm max}_{S_3})$.
Recall that both $m^{\rm max}_{S_2}$ and $m^{\rm max}_{S_3}$ are
functions of $m_{S_1}$.

It is quite clear that each of the three production cross sections
will exhibit its own minimum (and maximum) for a certain value of
$m_{S_1}$ in the range between its lower and upper bound.
Naturally, $\sigma_1 (R_1, R_2, m_{S_1})$ is minimized when
$m_{S_1}$ is maximized.
That is, the minimum of $\sigma_1 (R_1, R_2, m_{S_1})$ is given by
$\sigma_1 (R_1, R_2, m^{\rm max}_{S_1})$.
Let us denote it by $\sigma^{\rm min}_1 (R_1, R_2)$.

For $\sigma_2 (R_1, R_2, m^{\rm max}_{S_2})$
and $\sigma_3 (R_1, R_2, m^{\rm max}_{S_3})$, it is not straightforward
to tell where their minima occur within the allowed range of $m_{S_1}$,
because both $m^{\rm max}_{S_2}$ and $m^{\rm max}_{S_3}$ are dependent
not only on $m_{S_1}$ but also $R_1$ and $R_2$.
Nontheless, the minima of $\sigma_2 (R_1, R_2, m^{\rm max}_{S_2})$
and $\sigma_3 (R_1, R_2, m^{\rm max}_{S_3})$ would exist for certain
values of $m_{S_1}$. Let them be denoted by $\sigma^{\rm min}_2 (R_1, R_2)$
and $\sigma^{\rm min}_3 (R_1, R_2)$, respectively.
(Note that the value of $m_{S_1}$ that yields $\sigma^{\rm min}_2 (R_1, R_2)$
is generally different from the one that does
$\sigma^{\rm min}_3 (R_1, R_2)$, for given $R_1$ and $R_2$.)

Then, by comparing the three minima with each other, we can establish
\begin{equation}
        \sigma (R_1, R_2) = \mbox{max} \{ \sigma^{\rm min}_1 (R_1, R_2),
                \sigma^{\rm min}_2 (R_1, R_2),
                \sigma^{\rm min}_3 (R_1, R_2) \}
\end{equation}
for given $R_1$ and $R_2$. In other words,
\[
        \sigma_i (R_1, R_2, m_{S_i}) \ge
        \sigma_i (R_1, R_2, m^{\rm max}_{S_i}) \ge
        \sigma^{\rm min}_i (R_1, R_2, m_{S_i})
        = \sigma (R_1, R_2)
\]
for some $i$.
The meaning of $\sigma(R_1, R_2)$ cannot be missed: At least one of
the three scalar Higgs bosons has its production cross section
via the Higgsstrahlung process in $e^+e^-$ collisions larger than
$\sigma(R_1, R_2)$ for a certain set of relevant parameters of the 
NMSSM.

The third step is to plot $\sigma(R_1, R_2)$ in the ($R_1$, $R_2$)-plane 
for given ${\sqrt s}$. The triangular area in the plane defined by
$0 \le R^2_1 \le 1$, $0 \le R^2_2 \le 1$,
and $0 \le (1 -R^2_1 -R^2_2) \le 1$ represents the whole space of the
relevant parameters of the NMSSM, because it is the entire physical area 
in the ($R_1$, $R_2$)-plane.

We find that $\sigma (R_1, R_2)$ vanishes for the collider energy $\sqrt{s}$
less than $E_T = m^{\rm max}_{S_1} +m_Z$. This is the case for LEP II 
with $\sqrt{s} \le 205$ GeV, which implies that even the highest proposed
energy of LEP II would most probably not be able to test our model
completely.
If ${\sqrt s}$ is larger than $E_T$, then $\sigma(R_1, R_2)$ never
vanishes for that collider energy.
Thus, for given ${\sqrt s} \ge E_T$, $\sigma(R_1, R_2)$ exhibits
a nonzero minimum somewhere in the ($R_1$, $R_2$)-plane.

The final step is to establish the minimum of $\sigma(R_1, R_2)$
in the ($R_1$, $R_2$)-plane for given ${\sqrt s}$.
Let it be denoted as $\sigma_0 ({\sqrt s})$. Thus, we have
\[
        \sigma(R_1, R_2) \ge \sigma_0 ({\sqrt s})
\]
This minimum is a universal minimum in the sense that it is independent
of the parameters of the NMSSM. It is the absolute lower bound
on at least one of $\sigma_i (R_1, R_2, m_{S_i}) \ge \sigma_0({\sqrt s})$
for some $i$: At least one of the three
scalar Higgs bosons may be produced via the Higgsstrahlung process
in $e^+e^-$ collisions with production cross section larger than
$\sigma_0({\sqrt s})$.
This minimum is therefore a characteristic quantity of the NMSSM.


Let us now plot $\sigma(R_1, R_2)$ in the ($R_1$, $R_2$)-plane for
given ${\sqrt s}$ in order to evaluate $\sigma_0({\sqrt s})$, its 
minimum.
For simplicity, we set $m^{\rm max}_{S_1} = 145$ GeV.
The dependence of $\sigma(R_1, R_2)$, and hence $\sigma_0({\sqrt s})$,
on $m^{\rm max}_{S_1}$ is quite small for ${\sqrt s} \ge 500$ GeV, as 
will
be shown shortly.~

In Fig. 3, we plot $\sigma(R_1, R_2)$ for ${\sqrt s} = 500$ GeV.
The minimum is found to be about 16 fb.~
When the discovery limit is about 30 events, one would need a luminosity 
of about 25 fb for the future $e^+e^-$ collider, which is a realistic 
one.
In Fig. 4 and Fig. 5, we repeat our plottings for ${\sqrt s} = 1000$
GeV and 2000 GeV, respectively.
We find that the minimum of the production cross section is
4 fb for ${\sqrt s} = 1000$ GeV and 1 fb for 2000 GeV.

To see the dependence of $\sigma_0({\sqrt s})$ on the collider energy,
we plot it as a function of ${\sqrt s}$ in Fig. 6 for some values
of $m^{\rm max}_{S_1}$.
It is observed that $\sigma_0({\sqrt s})$ becomes largest for around
${\sqrt s} = 300$ GeV.
Moreover, the effect of $m^{\rm max}_{S_1}$ is most dominant there.
For ${\sqrt s} \ge 500$ GeV, $\sigma_0({\sqrt s})$ decreases
as ${\sqrt s}$ increases. Also, the effect of $m^{\rm max}_{S_1}$
on $\sigma_0({\sqrt s})$ becomes rather small there.

As an illustration, the individual production cross sections
$\sigma_i (R_1, R_2, m_{S_i})$ are calculated for an exemplary set
of the relevant parameters of the NMSSM.
Here, not only the Higgsstrahlung process
$e^+e^- \rightarrow S_i Z \rightarrow S_i b{\bar b}$ but also the
contributions from two other important processes are considered.
The two other processes are the process
$e^+e^- \rightarrow Z \rightarrow b{\bar b} \rightarrow S_i b{\bar b}$
where $S_i$ is radiated off from $b$ or ${\bar b}$, and the process
$e^+e^- \rightarrow Z \rightarrow S_i P_j \rightarrow S_i b{\bar b}$
where $P_j$ ($j$ = 1,2) is a pseudoscalar Higgs boson.

The results are plotted in Fig. 7 and Fig. 8, as functions of the
collider energy ${\sqrt s}$, for $A_{\lambda} = 220$ GeV,
$A_k = 160$ GeV, $x = 1000$ GeV, $\lambda = 0.12$, $k = 0.04$,
and $\tan\beta = 2$.
The production cross sections in Fig. 7 are the tree-level ones, whereas
those in Fig. 8 are those with the 1-loop corrections via the effective
potential. The relevant parameters for the 1-loop corrections are
taken as $A_t = 0$, $m_{\tilde t} = 1000$ GeV, and $m_t = 175$ =
GeV.

For this exemplary set of parameters, the 1-loop corrections is seen to
be important for the collider energy of about 150 GeV and decreases
as the collider energy increases.
Also, for ${\sqrt s} = 500$ GeV and beyond, we find that $S_2$ would
most dominantly be produced for the exemplary set of parameters,
both at the tree level and the 1-loop level.


The possibility of detecting one of the three scalar Higgs bosons in
the NMSSM at the future $e^+e^-$ linear collider has been analyzed
by examining their productions via the Higgsstrahlung process.
The whole space of the relevant parameters of the NMSSM is
conveniently represented by a triangular area in the $(R_1, R_2)$-plane, 
where the production cross sections of the three scalar Higgs bosons
in $e^+e^-$ collisions are evaluated for given c.m. energy of
the $e^+e^-$ collider.

For the three production cross sections, we have searched their minima
in the $(R_1, R_2)$-plane.
Then, from the three minima of the production cross sections, we have
established a characteristic quantity for them, $\sigma_0 ({\sqrt s})$,
which is a kind of universal minimum of the production cross section.
Actually, it is the largest one among the three minima of the
production cross sections.
We have obtained that $\sigma_0 ({\sqrt s}) = 16$ fb, 4 fb, and 1 fb
for ${\sqrt s} = 500$ GeV, 1000 GeV, and 2000 GeV, respectively.
As stressed in the preceding section, these numbers indicate that
at least one of the three scalar Higgs bosons may be produced via
the Higgsstrahlung process in $e^+e^-$ collisions with production
cross section larger than $\sigma_0({\sqrt s})$.

What we have found indicate that, for a certain set or regions of
the relevant parameters of the NMSSM, at least one of the three
scalar Higgs bosons might have a reasonable mass and a reasonable
production cross sections such that it might be
produced at the future $e^+e^-$ colliders.
We conclude that the Higgs sector of the NMSSM can most probably
be tested at the future linear $e^+e^-$ colliders with ${\sqrt s} =
500$, 1000, or 2000 GeV.

After we completed this article, we have been informed that a similar
conclusion has been drawn by Gunion, Haber, and Moroi [14]. However, 
their approach is quite different from ours.

\vfil\eject

This work is supported in part by the joint program between DFG and 
KOSEF
and in part by the Basic Science Research Institute Program,
Ministry of Education, BSRI-97-2442 (Korea).

\vfil\eject


\vfil\eject

\noindent
{\bf Figure Captions}

\vskip 0.3 in
\noindent
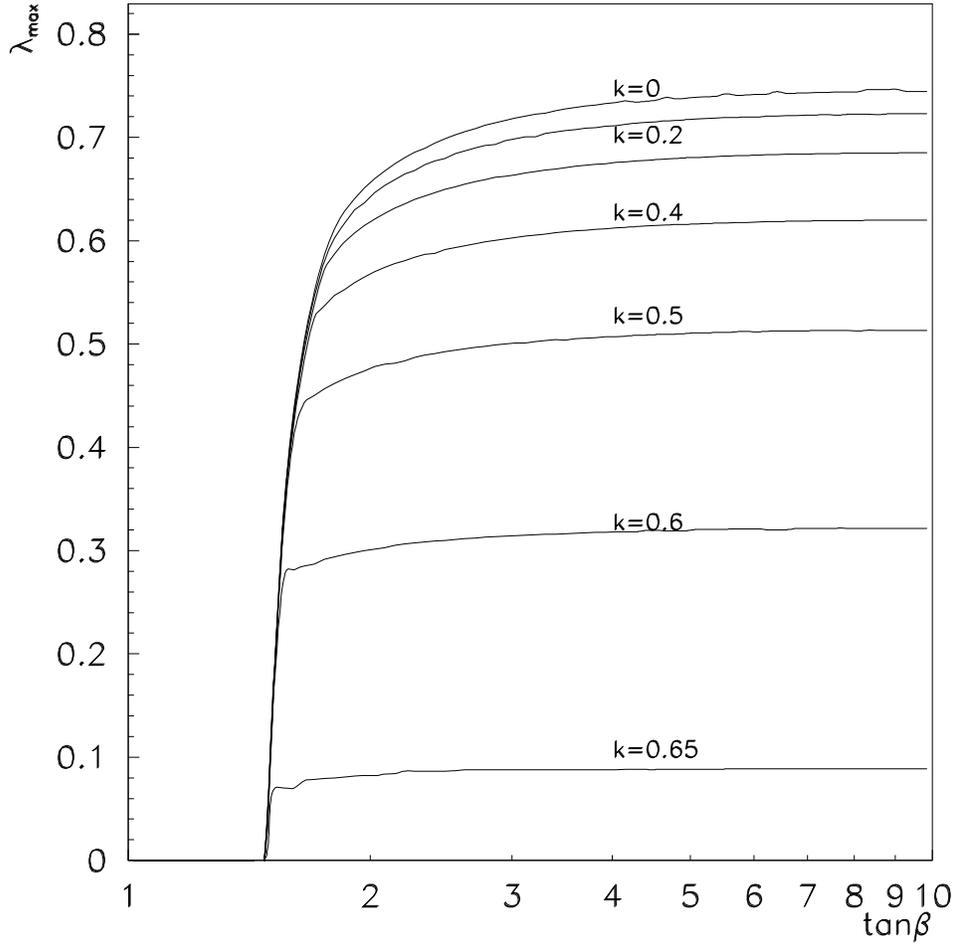
Figure 1:\, The upper bound on $\lambda$ as function of $\tan\beta$
for some values of $k$, for $m_t = 175$ GeV. 

\noindent
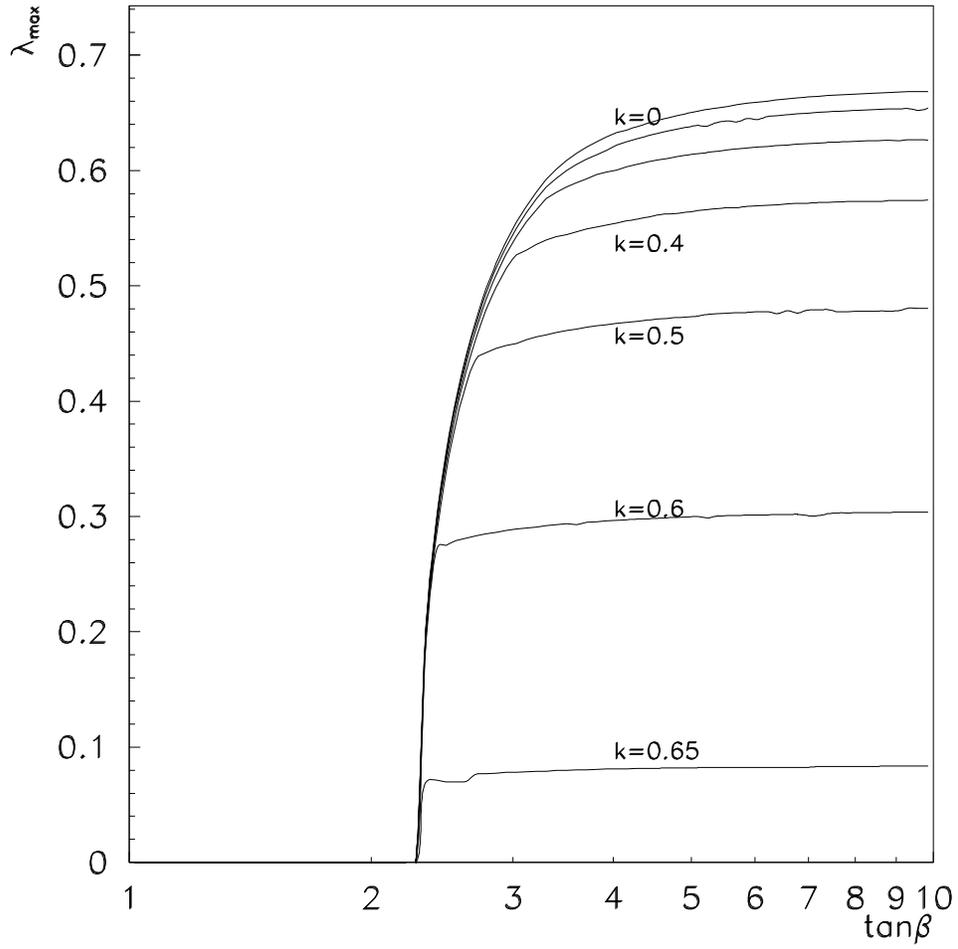
Figure 2:\, The upper bound on $\lambda$ as function of $\tan\beta$
for some values of $k$, for $m_t = 190$ GeV.

\noindent
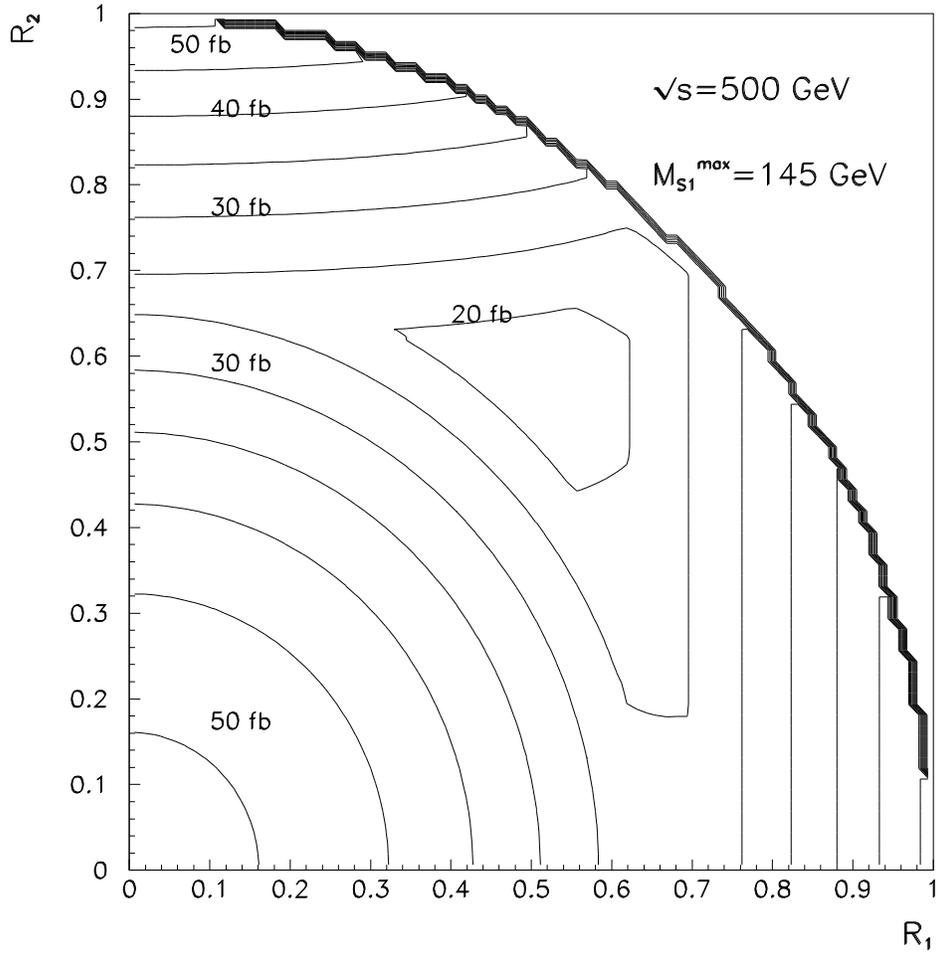
Figure 3:\, The contours of $\sigma(R_1, R_2)$ in the $(R_1, R_2)$-plane
for $m^{\rm max}_{S_1} = 145$ GeV and ${\sqrt s} = 500$ GeV.

\noindent
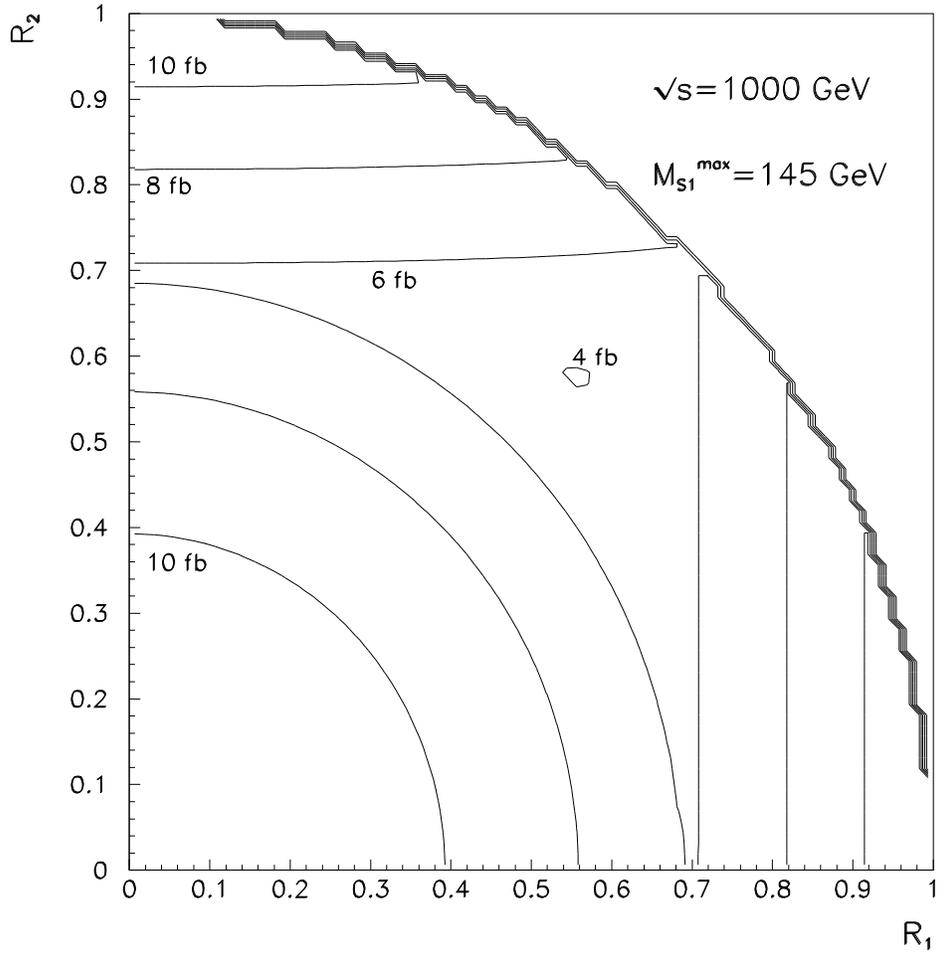
Figure 4:\, The contours of $\sigma(R_1, R_2)$ in the $(R_1, R_2)$-plane
for $m^{\rm max}_{S_1} = 145$ GeV and ${\sqrt s} = 1000$ GeV.

\noindent
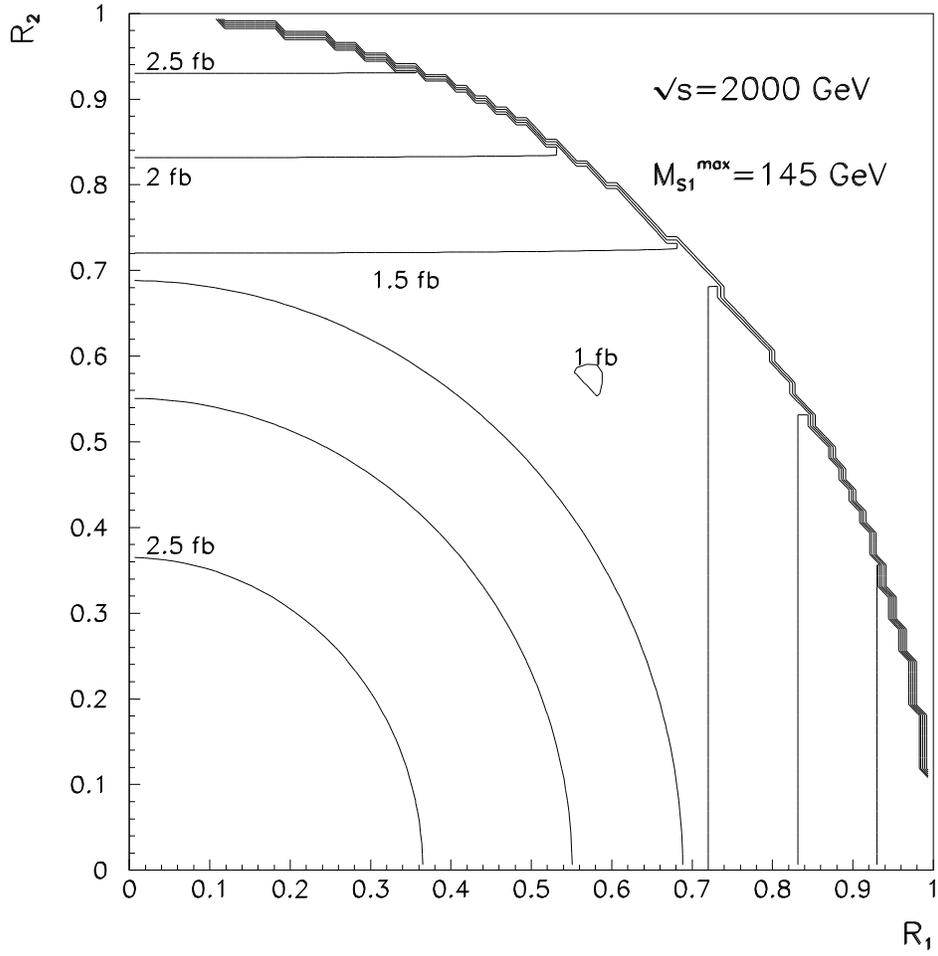
Figure 5:\, The contours of $\sigma(R_1, R_2)$ in the $(R_1, R_2)$-plane
for $m^{\rm max}_{S_1} = 145$ GeV and ${\sqrt s} = 2000$ GeV.

\noindent
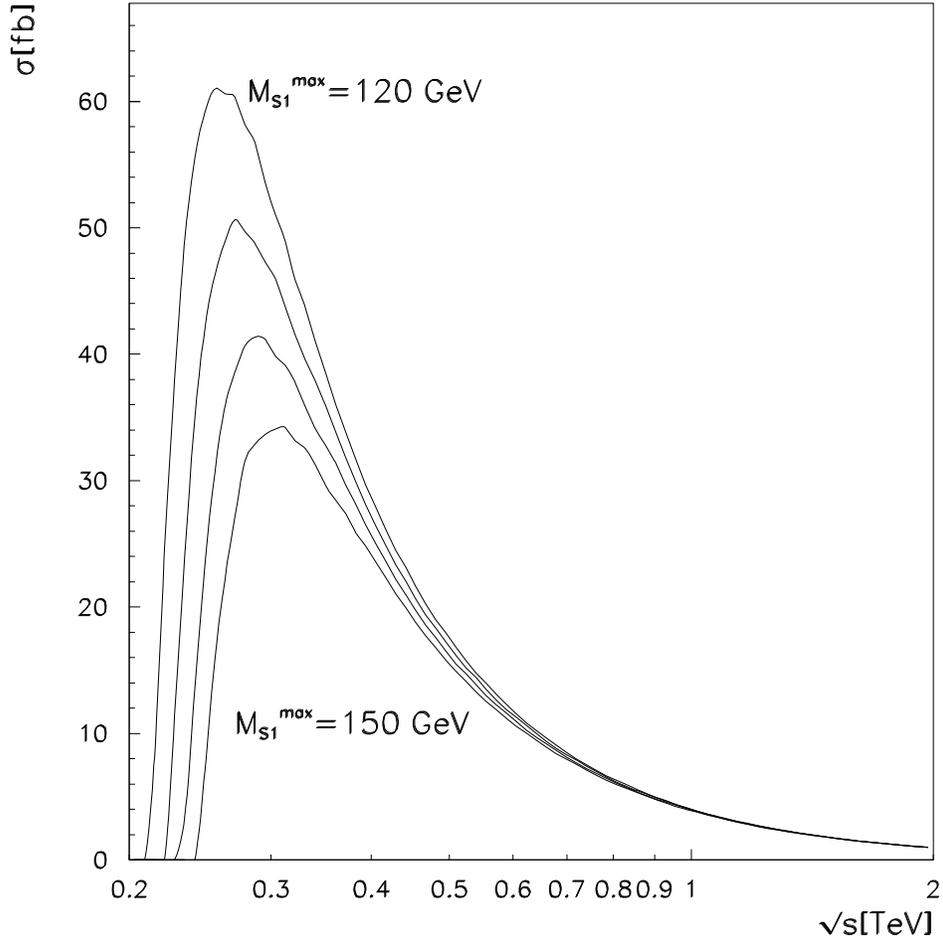
Figure 6:\, $\sigma_0 ({\sqrt s})$, the minimum of $\sigma(R_1, R_2)$,
as function of the collider energy ${\sqrt s}$ for some
values of $m^{\rm max}_{S_1}$.

\noindent
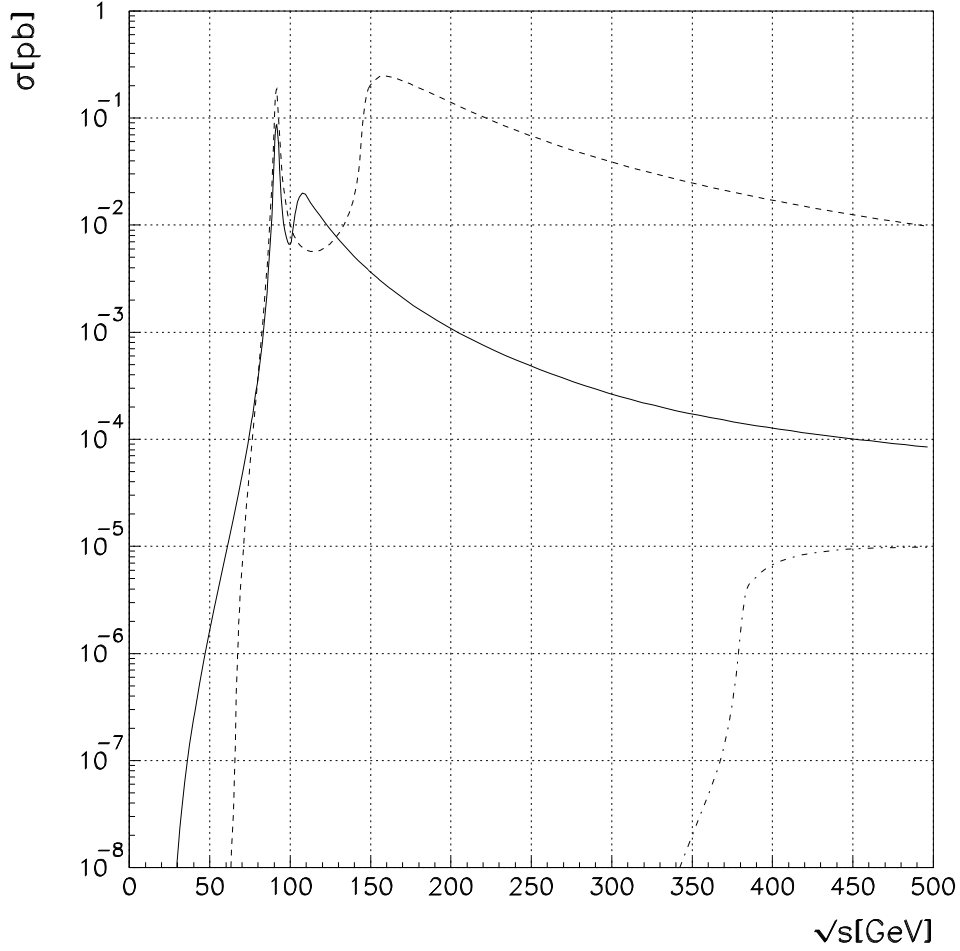
Figure 7:\, The production cross sections at the tree level
$\sigma(e^+e^- \rightarrow b{\bar b}S_i)$ ($i = 1,2,3$) as function
of the collider energy ${\sqrt s}$, for $A_{\lambda} = 220$ GeV,
$A_k = 160$ GeV, $x = 1000$ GeV, $\lambda = 0.12$, $k = 0.04$,
and $\tan\beta = 2$.
The solid, dashed, and dashed-dotted curves correspond respectively to
the $S_1$, $S_2$, and $S_3$ productions.

\noindent
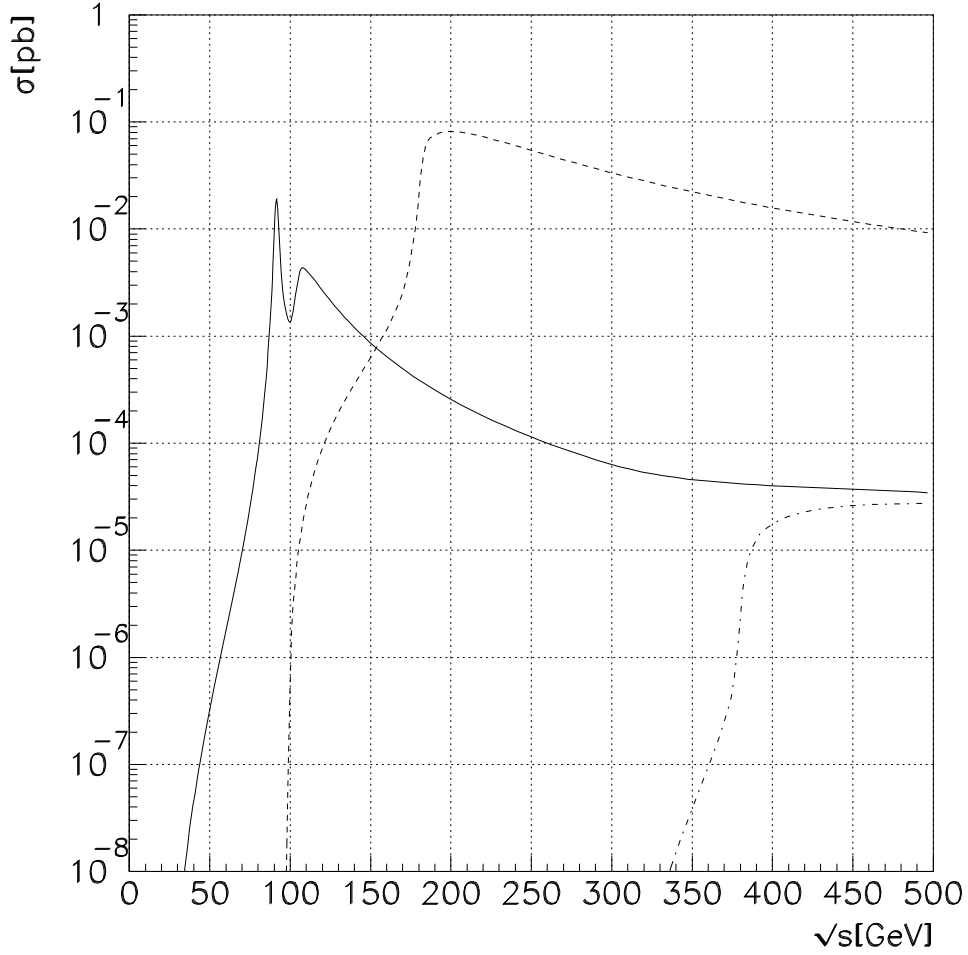
Figure 8:\, The production cross sections at the 1-loop level
$\sigma(e^+e^- \rightarrow b{\bar b}S_i)$ ($i = 1,2,3$) as function
of the collider energy ${\sqrt s}$, for $A_{\lambda} = 220$ GeV,
$A_k = 160$ GeV, $x = 1000$ GeV, $\lambda = 0.12$, $k = 0.04$,
and $\tan\beta = 2$. The 1-loop parameters are taken as
$A_t = 0$, $m_{\tilde t} = 1000$ GeV, and $m_t = 175$ GeV.
The solid, dashed, and dashed-dotted curves correspond respectively to
the $S_1$, $S_2$, and $S_3$ productions.

\vfil\eject

\end{document}